\documentclass[%
 reprint,
 superscriptaddress,
 amsmath,amssymb,
 aps,
 pra,
 longbibliography,
]{revtex4-2}

\usepackage{soul}     
\usepackage{graphicx} 
\usepackage{dcolumn}  
\usepackage{bm}       
\usepackage{color}
\usepackage[colorlinks,urlcolor=blue,citecolor=blue,linkcolor=blue,pdfstartview=FitH]{hyperref}
\usepackage{enumerate}
\usepackage{amsmath}
\usepackage{cases}
\usepackage{microtype}
\usepackage{siunitx}
\usepackage{subdepth}
\usepackage{todonotes}
\usepackage[utf8]{inputenc}
\usepackage[T1]{fontenc}
\usepackage{txfonts}           

\newcommand{\diff}{\mathrm{d}}
\newcommand{\Lc}{L_\mathrm{c}}
\newcommand{\nv}{n_\mathrm{v}}
\newcommand{\umean}{\bar{\rm u}_{\rm v}}
\newcommand{\mob}{\mu_{\rm v}}

\begin{document}

\title{Crossover in the dynamical critical exponent of a quenched two-dimensional Bose gas}

\author{A. J. Groszek}
\affiliation{Joint Quantum Centre (JQC) Durham--Newcastle, School of Mathematics, Statistics and Physics, Newcastle University, Newcastle upon Tyne, NE1 7RU, United Kingdom}
\author{P. Comaron}
\affiliation{Joint Quantum Centre (JQC) Durham--Newcastle, School of Mathematics, Statistics and Physics, Newcastle University, Newcastle upon Tyne, NE1 7RU, United Kingdom}
\affiliation{Institute of Physics, Polish Academy of Sciences, Al. Lotnik\'ow 32/46, 02-668 Warsaw, Poland}
\author{N. P. Proukakis}
\affiliation{Joint Quantum Centre (JQC) Durham--Newcastle, School of Mathematics, Statistics and Physics, Newcastle University, Newcastle upon Tyne, NE1 7RU, United Kingdom}
\author{T. P. Billam}
\affiliation{Joint Quantum Centre (JQC) Durham--Newcastle, School of Mathematics, Statistics and Physics, Newcastle University, Newcastle upon Tyne, NE1 7RU, United Kingdom}

\begin{abstract}
We study the phase ordering dynamics of a uniform Bose gas in two dimensions
following a quench into the ordered phase. We explore the crossover between dissipative and conservative evolution by performing numerical simulations within the classical field methodology. Regardless of the dissipation strength, we find clear evidence for universal scaling, with dynamical critical exponent $z$ characterising the growth of the correlation length. In the dissipative limit we find growth consistent with the logarithmically corrected law $[t/\log(t/t_0)]^{1/z}$, and exponent $z=2$, in agreement with previous studies. Decreasing the dissipation towards the conservative limit, we find strong numerical evidence for the expected growth law $t^{1/z}$. However, we observe a smooth crossover in $z$ that converges to an anomalous value distinctly lower than $2$ at a small finite dissipation strength. We show that this lower exponent may be attributable to a power-law vortex mobility arising from vortex--sound interactions.
\end{abstract}

\maketitle

\section{Introduction}

A many-body system quenched from a disordered to an ordered phase has long been a topic of interest in nonequilibrium physics. Following the quench the system relaxes toward a new equilibrium configuration via a process of domain coarsening, with an associated growth of the correlation length $\Lc(t)$. The dynamical scaling hypothesis posits that at sufficiently late times the system should approach a statistically invariant state in which $\Lc$ becomes the only relevant length scale. In this state, the correlation length is predicted to grow $\sim t^{1/z}$, where $z$ is the dynamical critical exponent~\cite{bray_theory_1994}. In the classical theory of phase ordering kinetics, coarsening is described in terms of the dynamics and annealing of topological defects, with conservation laws and dimensionality playing a key role in determining $z$~\cite{bray_theory_1994}. Extensive numerical studies in two-dimensional (2D) systems such as Ising~\cite{humayun_non-equilibrium_1991, kim_nonequilibrium_2003} and XY~\cite{rojas_dynamical_1999, bray_breakdown_2000, jelic_quench_2011} models have provided broad support for this simple physical picture.

In recent years, ultracold Bose gases have become an established platform for the exploration of nonequilibrium dynamics in a quantum setting. Owing to their exquisite tunability, experiments have been able to use these gases to probe physics such as the Kibble--Zurek mechanism~\cite{weiler_spontaneous_2008, lamporesi_spontaneous_2013, chomaz_emergence_2015, navon_critical_2015} and quantum turbulence~\cite{navon_emergence_2016, navon_synthetic_2019, gauthier_giant_2019, johnstone_evolution_2019}, as well as scale invariant dynamics following a quench, similar to the scenario described above~\cite{erne_universal_2018, prufer_observation_2018, eigen_universal_2018, glidden_bidirectional_2021}. In this last context, the concept of nonthermal fixed points~\cite{berges_nonthermal_2008, berges_nonthermal_2009, scheppach_matter_2010, nowak_superfluid_2011, orioli_universal_2015, karl_strongly_2017, schachner_universal_2017} has emerged as a powerful theoretical description of scaling behaviour in which topological defects appear to play a less crucial role than in phase ordering kinetics~\cite{schachner_universal_2017}.

Theoretical studies have addressed coarsening following an instantaneous quench in 2D bosonic systems such as binary~\cite{hofmann_coarsening_2014} and spinor~\cite{williamson_universal_2016, williamson_coarsening_2017} condensates and driven--dissipative systems~\cite{kulczykowski_phase_2017, comaron_dynamical_2018, gladilin_multivortex_2019, mei_spatiotemporal_2021}. However, in the apparently simple case of a quenched scalar 2D Bose gas there remain open questions regarding the link between coarsening behaviour and conservation laws in the dynamics. These stretch back to the well-known classification of dynamical universality classes established in Ref.~\cite{hohenberg_theory_1977}. What the precise value of $z$ is for conservative dynamics, and how $z$ varies with the dissipation strength, remain important open questions. These questions are of particular relevance in Bose gases, where there exist both conservative and nonconservative classical field descriptions of the dynamics~\cite{blakie_dynamics_2008}. The former conserve energy and particle number~\cite{damle_phase_1996, davis_simulations_2001}. The latter include dissipation~\cite{stoof_coherent_1999, stoof_dynamics_2001, gardiner_stochastic_2002, gardiner_stochastic_2003, bradley_bose-einstein_2008}; they have no conserved quantities, and in the dissipative limit they reduce to a purely relaxational time-dependent Ginzburg--Landau equation. Hence, for nonconservative dynamics the relevant dynamical universality class would appear to be Model A \cite{hohenberg_theory_1977}. There is general theoretical and numerical agreement that for Model A, $z=2$ with logarithmic corrections \cite{halperin_calculation_1972, yurke_coarsening_1993, rutenberg_energy-scaling_1995, bray_breakdown_2000, jelic_quench_2011}. However, as noted in Ref.~\cite{hohenberg_theory_1977}, the coarsening behaviour of a Bose gas with conservative dynamics is theoretically less tractable. Previous numerical studies in this scenario measured contradictory exponents $z \sim 1$~\cite{damle_phase_1996} (with no quoted uncertainty) and $z=1.8(3)$~\cite{karl_strongly_2017, karl_uncertainty_note}; meanwhile, Refs.~\cite{koo_coarsening_2006, nam_coarsening_2012} studied the related conservative XY model and proposed that $z=2$ but with a different form of logarithmic corrections to Model A. Simulations of coarsening in a nonconservative Bose gas have yielded exponents $z = 2.0(2)$~\cite{comaron_quench_2019} and $z = 1.9(2)$~\cite{karl_strongly_2017, karl_uncertainty_note}, although the weak dissipation included in such works places these results between the purely dissipative Model A and the conservative limit. Additionally, while Refs.~\cite{damle_phase_1996, karl_strongly_2017, comaron_quench_2019} did not include logarithmic corrections, they also did not rule out their relevance. As such, an overall picture remains elusive.

Here, we revisit the problem of coarsening in a 2D scalar Bose gas after an instantaneous quench. We apply two classical field methods---the (conservative) projected Gross--Pitaevskii equation (PGPE) and the (nonconservative) stochastic projected Gross--Pitaevskii equation (SPGPE)---and explore the crossover between the conservative and fully dissipative (Model A) limits. Our large scale simulations, large ensemble sizes, and  careful analysis of fitting and   systematic uncertainties allow us to tightly constrain the exponent values, yielding strong evidence of a crossover in the value of $z$ between the dissipative and conservative limits. In the dissipative limit, we find an exponent consistent with $z=2$ with logarithmic corrections, in good agreement with previous results for Model A. For decreasing dissipation, we find that the exponent decreases to $z \approx 1.7$ in the conservative limit for the parameters explored here. We analyse the vortex motion and find that the decrease in $z$ may be attributable to a power-law vortex mobility resulting from vortex--sound interactions.

\section{Simulations \label{sec:simulations}}

To describe a Bose gas at finite temperature, we adopt a
classical field model~\cite{blakie_dynamics_2008},
\begin{equation} \label{eq:SPGPE}
    \diff \psi = \mathcal{P} \biggl \lbrace - i\frac{\alpha}{\hbar} L_{\rm GP} \psi \diff t + \frac{\gamma}{\hbar} ( \mu - L_{\rm GP}) \psi \diff t + \diff W \biggr \rbrace,
\end{equation}
where $L_{\rm GP} = - (\hbar^2 / 2m) \nabla^2 + g |\psi| ^2$. In this model, the gas is represented with a complex scalar field $\psi(\textbf{r}, t)$, which includes contributions from all highly occupied single-particle modes of the system, up to some chosen cutoff in the single-particle energy spectrum. The gas is considered to be in contact with a thermal reservoir at temperature $T$ and with chemical potential $\mu$ (corresponding to the above-cutoff atoms), with which it can exchange both energy and particles. The projection operator $\mathcal{P}$ ensures that no population is transferred outside the chosen subset of single-particle modes during the evolution, while the constants $m$ and $g$ correspond to the particle mass and the 2D interaction strength, respectively. The dimensionless effective dissipation rate $\gamma$ controls the strength of the coupling between the system and the bath, and $\diff W(\textbf{r},t)$ is a complex Gaussian noise term satisfying $\langle \diff W^* (\textbf{r},t) \diff W (\textbf{r}',t) \rangle = (2 \gamma k_{\rm B} T / \hbar) \delta (\textbf{r} - \textbf{r}') \diff t$. With $\alpha=1$, this model is known generally as the SPGPE \footnote{The SPGPE was established in Refs.~\cite{gardiner_stochastic_2002, gardiner_stochastic_2003, bradley_bose-einstein_2008}. In the terminology of Ref.~\cite{blakie_dynamics_2008} we use the \textit{simple growth SPGPE}. Non-projected stochastic Gross--Pitaevskii equations have also been developed~\cite{stoof_coherent_1999, stoof_dynamics_2001}.}. The rate $\gamma$ in the SPGPE can be predicted \textit{a     priori}~\cite{bradley_bose-einstein_2008, blakie_dynamics_2008} in (near-) equilibrium   situations~\cite{rooney_persistent-current_2013}. Far from equilibrium,   quantitative agreement with experiments is improved by treating $\gamma$ as a   free parameter, typically with $\gamma \lesssim   0.02$~\cite{weiler_spontaneous_2008, rooney_persistent-current_2013, ota_collisionless_2018, liu_dynamical_2018}. Here, by varying $\gamma$ freely we explore the range applicable to experiments and also eludicate the conservative and dissipative (Model A) limits. In the limit $\gamma \to 0$, the coupling is removed, and Eq.~\eqref{eq:SPGPE} reduces to the PGPE, for which both the energy $E = \int ( \hbar^2 |\nabla \psi|^2 / 2m + g | \psi |^4 / 2 ) \diff \mathbf{r}$ and norm $N = \int | \psi |^2 \diff \mathbf{r}$ are conserved under time evolution. For $\gamma \gg 1$, on the other hand, the first term on the right hand side of Eq.~\eqref{eq:SPGPE} becomes negligible, and the dissipative Model A is recovered. In practice, we set $\alpha=0$ to access Model A, and use $\alpha=1$ otherwise.

In this work, we consider a system in a doubly periodic square domain of size $L \times L$. The properties of the thermal bath ($\mu$ and $T$) are held fixed, and only the dissipation rate $\gamma$ is varied. Two quench protocols are used, depending on the choice of $\gamma$. For $\gamma>0$ we begin with $\psi = 0$ and instantaneously switch on the reservoir coupling at time $t=0$, forcing the classical field density to grow nonadiabatically. For $\gamma=0$, we populate a disk of modes in wavenumber space uniformly and with random phase, while constraining the energy- and particle-densities to both match the $\gamma>0$ simulations at late times (see Appendix~\ref{app:ICs} for further details). Both types of quench initialise the system far from equilibrium, with a high density of quantised vortices and antivortices.

The single-particle modes for our chosen geometry are plane waves satisfying $|\textbf{k}| < k_{\rm cut}$ for some wavenumber cutoff $k_{\rm cut}$. To prevent aliasing, the wavenumber cutoff is set to $k_{\rm cut} = \pi / (2 \Delta x)$ (half the Nyquist wavenumber of the grid), with a numerical grid spacing of $\Delta x \approx 0.7 \, \xi$, where $\xi = \hbar / (m \mu)^{1/2}$ is the healing length. We set $T \approx 2.1 \, \mu / k_{\rm B}$ and $g \approx 0.17 \, \hbar^2 / m$ in Eq.~\eqref{eq:SPGPE}, resulting in an occupation of $\sim 1$ particle per mode at the cutoff~\cite{blakie_dynamics_2008}. This corresponds to a quench deep into the ordered phase. Quenches for other parameters, as well as quenches between two temperatures within the ordered phase~\cite{yurke_coarsening_1993, bray_breakdown_2000, forrester_exact_2013}, are an interesting avenue for future work. The $\gamma>0$ SPGPE is solved numerically using XMDS2~\cite{dennis_xmds2:_2013}, while the $\gamma=0$ PGPE is parallelised on nVidia Tesla V100 GPUs using CUDA \cite{cuda_toolkit}, allowing for the significantly longer evolution time required in this limit. The system and ensemble sizes are respectively chosen to be $L \approx 262 \, \xi$ and $\mathcal{N}=400$ for $\gamma > 0.1$, and $L \approx 363 \, \xi$ and $\mathcal{N}=256$ for $\gamma \leq 0.1$.

\begin{figure}[t]
    \centering
    \includegraphics[width=\columnwidth]{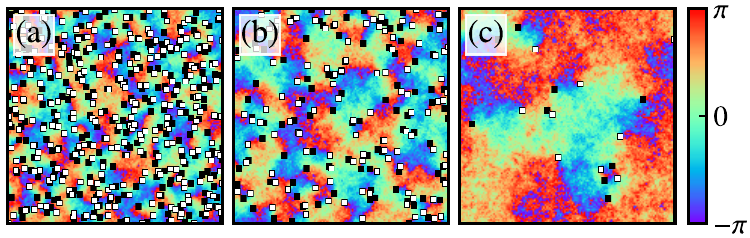}
    \caption{Evolution of the phase of the field $\psi$ in the $\gamma = 0$ system. Panels (a)--(c) correspond to times $\mu t / \hbar \approx \lbrace 200, 2000, 20\,000 \rbrace$, respectively. White (black) squares indicate the locations of vortices (antivortices). See Appendix~\ref{app:movies} and the Supplemental Material~\cite{supplement} for movies of the evolution.}
    \label{fig:phase_vortices}
\end{figure}

\begin{figure*}[t]
    \centering
    \includegraphics[width=2\columnwidth]{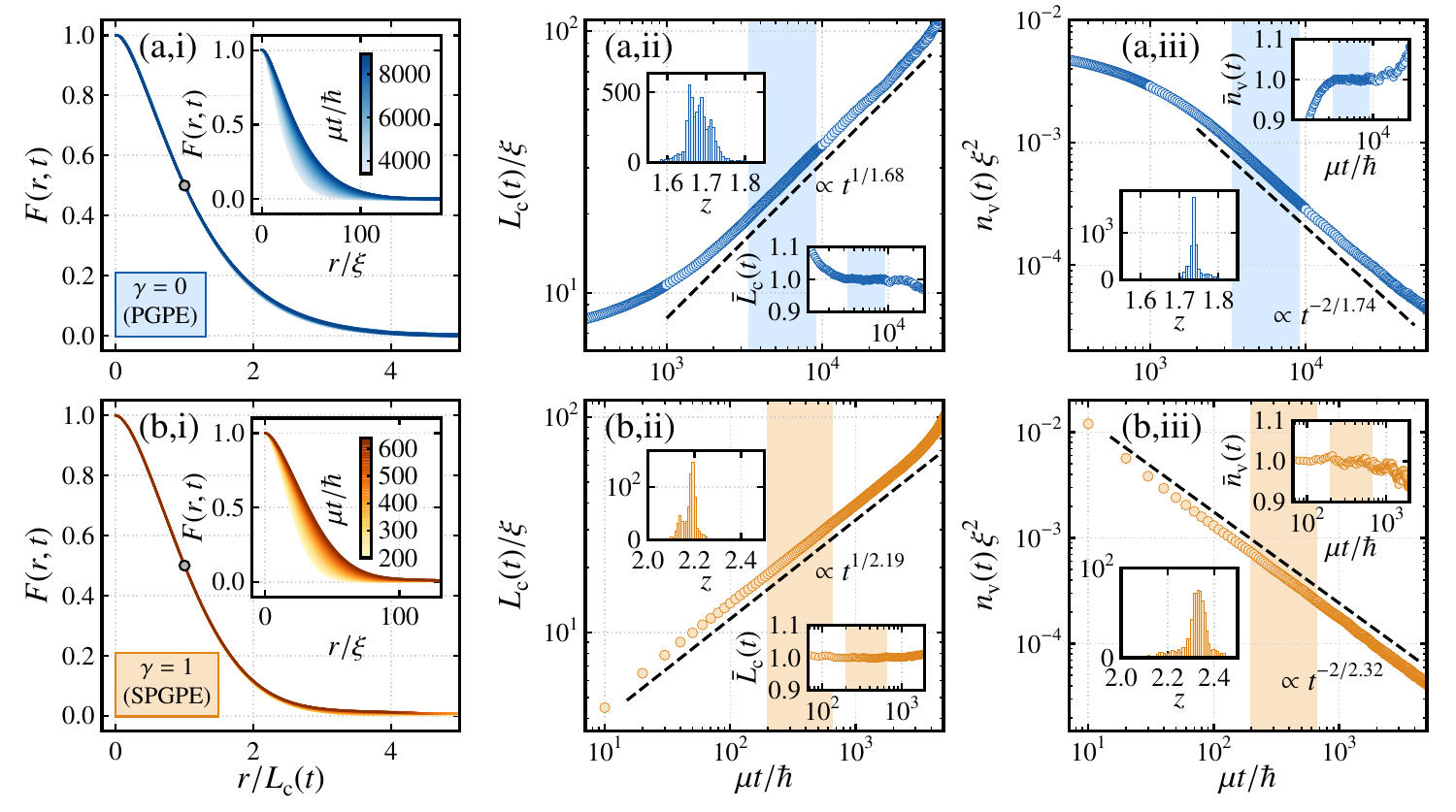}
    \caption{Scaling behaviour with $\alpha=1$, and dissipation rates (a) $\gamma=0$ and (b) $\gamma=1$. Ensemble averaged scaling function $F(r,t)$ plotted against radial distance $r$ (i), both before (inset) and after (main frame) rescaling by the correlation length $\Lc(t)$. The time at which each curve has been sampled is denoted by the colorbar in the insets, and a grey dot signifies the threshold $F_0=0.5$ used to define $\Lc$. Evolution of the mean correlation length $\Lc(t)$ (ii) and vortex density $\nv(t)$ (iii). In columns (ii,iii), the chosen scaling region is highlighted, and the best power-law fit to the data within that region is shown as a black dashed line (offset for visibility). In these four panels, the left insets are histograms showing the distribution of exponents $z$ measured from fits to subsets of the data within the highlighted region, while the right inset shows the compensated correlation length (ii) and vortex density (iii) as a function of time (horizontal axis same as for main frames).}
    \label{fig:scaling_data}
\end{figure*}

\section{Analysis \label{sec:analysis}}
Following the quench, the Bose gas begins to relax and the vortex number decreases via the annihilation of vortex--antivortex pairs. This process is illustrated in Fig.~\ref{fig:phase_vortices}, where the phase, $\mathrm{arg} \lbrace \psi \rbrace$, is shown at three times during the evolution of the PGPE. The vortex density is seen to decrease over time, allowing regions of phase coherence to develop.

As a measure of the spatial coherence of the field at a given time, we calculate the first-order correlation function,
\begin{equation} \label{eq:correlation_function}
G(\mathbf{r},t) = \frac{\langle \psi^*(\mathbf{r}+\mathbf{r}', t) \psi(\mathbf{r}', t)  \rangle}{\sqrt{ \langle | \psi(\mathbf{r}+\mathbf{r}', t) |^2 \rangle \langle | \psi(\mathbf{r}', t) | ^2 \rangle}}.
\end{equation}
The angular brackets in this expression correspond to an average over both stochastic realisations and the co-ordinate $\mathbf{r}'$. The scaling hypothesis asserts that a universal form for this correlator, $G(r, t) = G_{\rm eq}(r) F(r, t)$, should emerge at late times~\cite{bray_theory_1994}. Here, $G_{\rm eq}(r) = G(r, t \to \infty)$ is the equilibrium correlation function, and $F$ is a scaling function that should have the form $F(r,t) = F(r/\Lc(t))$, with $F(0)=1$. The correlation length $\Lc(t)$ in this expression is defined as the average distance over which equilibrium correlations have been established at time $t$, corresponding to the average size of the phase domains.

We calculate $F(r,t)$ from our simulations by measuring both $G(r,t)$ and
$G_{\rm eq}(r)$, where the latter is obtained from a temporal and ensemble
average of $G(r,t)$ once the system has equilibrated (equilibration is inferred
from the stabilisation of the $k=0$ mode population). Below the
Berezinskii--Kosterlitz--Thouless transition~\cite{berezinskii_destruction_1971,
  berezinskii_destruction_1972, kosterlitz_ordering_1973}, we expect that
$G_{\rm eq}(r) \sim r^{-\eta}$ for $r \gg \xi$, where $0 \leq \eta(T) \leq 0.25$
is a temperature-dependent exponent~\cite{pethick_bose-einstein_2008}. We find
that $\eta \approx 0.06$ for our parameters, using a method described in
Ref.~\cite{nazarenko_bose-einstein_2014}. We measure the correlation length
$\Lc(t)$ as the radial distance satisfying $F(\Lc,t) = F_0$, where we set
$F_0=0.5$ (Appendix~\ref{app:systematics} provides further details regarding the choice of threshold). This choice assists in excluding discretisation
effects at small scales and finite-size effects at large scales. Similarly, we
should extract $\Lc(t)$ over a scaling window in time that both suppresses
finite-size effects at long times and excludes initial transients. In practice
we end the window as soon as $\Lc(t) > L/4$ in \textit{any one} of the
simulations in the ensemble; this stringent condition generally corresponds to
an \textit{average} $\Lc(t)$ of $\sim L /10$ at the end of the window. We start
the window as early as possible while ensuring that there are minimal deviations
from the unique scaling function $F(r/\Lc(t))$ (we quantify this in Appendix~\ref{app:DeltaF}).

\section{Results \label{sec:results}}

\subsection{Evidence of universal scaling \label{sub:scaling}}

The evolution of $F(r,t)$ is displayed in Fig.~\ref{fig:scaling_data}(a,i) and (b,i) for $\gamma=0$ and $\gamma=1$, respectively, and a collapse of the data onto a single unique curve is evident over the time windows shown. In Fig.~\ref{fig:scaling_data}(a,ii) and (b,ii), the evolution of $\Lc(t)$ is shown for the same two $\gamma$ values, and in both cases the data are well described by $\Lc(t) \sim t^{1/z}$ within the highlighted scaling window. To measure the exponent $z$ in each case, we fit a power-law to the data across all possible subintervals of $\geq 8$ consecutive points within the scaling window. This yields a distribution of $z$ values characterising the statistical uncertainty associated with temporal variations in the scaling of $\Lc(t)$ (left inset of each frame). We measure $z$ to be the mean of this distribution and estimate its statistical uncertainty to be the standard deviation. For these two cases, we obtain $z=1.68(4)$ ($\gamma=0$) and $z=2.19(3)$ ($\gamma=1$).  As an illustration of the goodness-of-fit, we also plot the compensated correlation length $\bar{L}_{\rm c}(t) = \Lc(t) / \Lc^{\rm fit}(t)$ [right insets of column (ii)]. Its value remains close to unity within the highlighted scaling window, and for some extent outside it.

Dynamical scaling is also expected to manifest in the decay of the vortex density $\nv(t)$~\cite{karl_strongly_2017, baggaley_decay_2018,  groszek_decaying_2020}, with $\nv(t) \sim \Lc^{-2}(t) \sim t^{-2/z}$ for randomly distributed defects. The mean vortex density is shown in Fig.~\ref{fig:scaling_data}(a,iii) and (b,iii), and a fit to the data within the highlighted window is found in the same way as for the $\Lc(t)$ curves. The left and right insets, as in column (ii), correspond respectively to the histogram of $z$ values from repeated fits, and the compensated mean vortex density $\bar{n}_{\rm v}(t) = \nv(t)/\nv^{\rm fit}(t)$. From this data, we measure $z=1.74(3)$ for $\gamma=0$ and $z=2.32(7)$ for $\gamma=1$; both of these values are slightly larger than those obtained from the corresponding fits to $\Lc(t)$. We note, however, that measuring $z$ from $\nv(t)$ is a less rigorous approach, because the vortices are not guaranteed to remain uniformly distributed. Indeed, within the scaling windows we measure negative nearest-neighbour vortex correlations, indicating a tendency toward dipole pairing of vortices and antivortices (see Appendix \ref{app:Cnn}). This effect is stronger for the larger $\gamma$.

\subsection{Measurements of the dynamical critical exponent \label{sub:z_measurements}}

\begin{figure}[b]
  \centering
  \includegraphics[width=1.0\columnwidth]{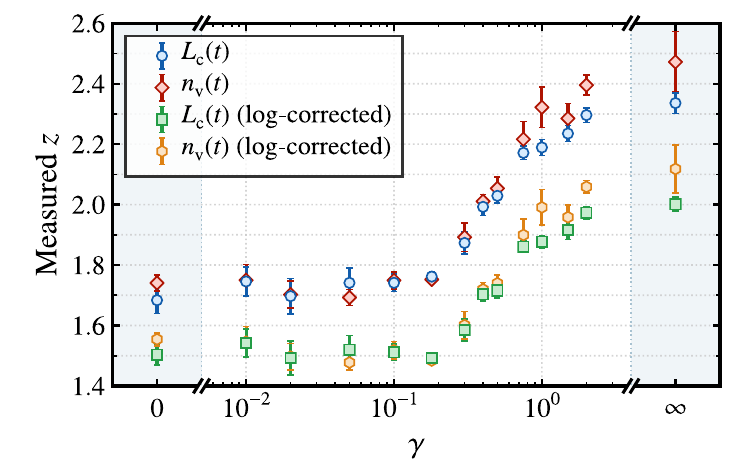}
  \caption{Measured dynamical critical exponent $z$ as a function of dimensionless dissipation rate $\gamma$. The error bars on each point correspond to the fitting uncertainty described in the text.}
  \label{fig:z_vs_gamma}
\end{figure}

We have repeated the analysis shown in Fig.~\ref{fig:scaling_data} for a range of $\gamma$ values, and find equally clear evidence for dynamical scaling in all cases. As above, the exponent $z$ is measured from fits to both $\Lc(t)$ and $\nv(t)$ in each case. We find a smooth crossover from $z \approx 2.3$ in Model A ($\gamma \to \infty$) to $z \approx 1.7$ in the PGPE ($\gamma=0$), as shown in Fig.~\ref{fig:z_vs_gamma}~\footnote{In fact, the SPGPE becomes independent of $\gamma$ for $\gamma \lesssim 0.1$, with $z \approx 1.7$ in this region.}. This clearly shows that the coarsening behaviour of the system changes as one crosses between Model A and conservative dynamics. We have performed several additional simulations and analyses to verify the robustness of this result (details are provided in Appendix~\ref{app:systematics}).

In the context of Model A dynamics, it has long been argued that $z=2$ with logarithmic corrections~\cite{yurke_coarsening_1993,  rutenberg_energy-scaling_1995}. These corrections are predicted to modify the scaling such that~\cite{bray_breakdown_2000, jelic_quench_2011, comaron_dynamical_2018, comaron_quench_2019}
\begin{equation} \label{eq:log_corrections}
\Lc \sim [t/ \log(t/t_0)]^{1/z},
\end{equation}
and hence $\Lc \sim t^{1/z}$ only in the limit $t \gg t_0$ for some microscopic timescale $t_0$. We find that a fit to Eq.~\eqref{eq:log_corrections} using our Model A $\Lc(t)$ data gives an exponent of $z=2$ if we choose $t_0 = 0.5 \, \hbar / \mu$~\footnote{Consistency with $z=2$ (to within our estimated uncertainties) is maintained for $0.3 \, \hbar / \mu \lesssim t_0 \lesssim 1.0 \, \hbar / \mu$.}, although we note that the fit quality is no better than an uncorrected power-law. Nonetheless, this establishes consistency between our results and the predicted behaviour for Model A. Log-corrected exponents fitted using the same $t_0$ at other values of $\gamma$ are shown in Fig.~\ref{fig:z_vs_gamma}, although existing arguments for this form of log-corrections only apply to Model A~\footnote{Outside the Model A limit we rescale $t_0$ by the length of the temporal unit vector in the complex plane: $t_0 \rightarrow 0.5 (\alpha^2 + \gamma^2)^{-1/2}\, \hbar / \mu$.}.

\subsection{Origin of $z<2$ \label{sub:zl2}}

To elucidate the origin of the measured crossover in $z$, we assemble movies of the evolution with vortex tracking (see Appendix~\ref{app:movies} and the Supplemental Material~\cite{supplement}). We observe a distinct qualitative change in the dynamics of vortices as $\gamma$ is varied. For $\gamma \ll 1$, nearest-neighbour vortex dipoles travel approximately perpendicular to their separation vector, and can traverse many times the average inter-defect distance before annihilating. By contrast, dipoles experience mutual attraction when $\gamma \gtrsim 1$, and hence vortices rarely travel beyond their closest neighbours before annihilating. We additionally observe that for $\gamma \gtrsim 1$ the vortices rapidly evolve to a state where mutual attraction becomes overwhelmed by fluctuations. It therefore seems possible that the crossover in $z$ may be explained in terms of vortex motion.

We start by assuming that: (i)~the correlation length grows at a rate determined by the mean vortex velocity $\umean \sim \diff \Lc / \diff t $; (ii)~the characteristic vortex velocity is determined by inter-vortex interactions and takes the form $\umean \sim \mob (\Lc) / \Lc$, where $\mob(\Lc)$ is the vortex mobility (see, e.g., Ref.~\cite{yurke_coarsening_1993}). We therefore must have
\begin{equation} \label{eq:dLdt_main}
\frac{\diff \Lc}{\diff t} \sim \frac{\mob(\Lc)}{\Lc},
\end{equation}
which can be integrated to yield $z$. A simple description of vortex motion in a Bose gas is a weakly damped point-vortex model, in which the vortices are taken to be point-particles with long-range interactions~\cite{weiss_nonergodicity_1991, billam_spectral_2015}. In this idealised case, $\mob$ is a constant by construction and therefore Eq.~\eqref{eq:dLdt_main} predicts $\Lc \sim t^{1/2}$, i.e.~$z=2$. If we instead assume a mobility $\mob \sim 1 / \log(\Lc/\xi)$ in Eq.~\eqref{eq:dLdt_main}, we arrive at Eq.~\eqref{eq:log_corrections} with $z=2$~\cite{yurke_coarsening_1993,rutenberg_energy-scaling_1995} and $t_0$ on the order of $\hbar / \mu$~\cite{bray_breakdown_2000, jelic_quench_2011}, which is in accordance with our fitting to Eq.~\eqref{eq:log_corrections} as described in the previous section.

\begin{figure}[t]
    \centering
    \includegraphics[width=\columnwidth]{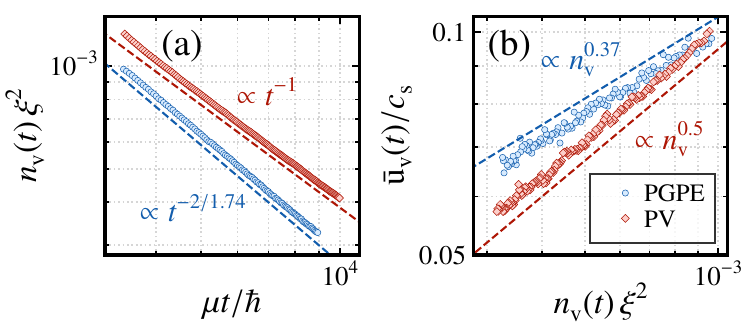}
    \caption{Comparison of vortex density evolution (a), and scaling of the mean vortex velocity (b), between the PGPE and the point-vortex (PV) model. The PGPE data are plotted only within the temporal scaling window identified in Fig.~\ref{fig:scaling_data}(a). Dashed lines show power-laws for comparison with the data. For improved visibility in (a), the PV vortex density data are offset by multiplying $\nv(t) \to 1.3\nv(t)$. The velocity is expressed in relation to the sound speed $c_{\rm s} = \mu \xi / \hbar$.}
    \label{fig:PVcomparison}
\end{figure}

A possible origin for $z < 2$ in the conservative limit is a power-law mobility $\mob \sim \Lc^\varepsilon$~\cite{pc_simula}, yielding $\Lc \sim t^{1/(2-\varepsilon)}$, i.e.
\begin{equation} \label{eq:z<2}
z=2-\varepsilon.
\end{equation}
Our PGPE ($\gamma=0$) data allow us to perform a measurement of the exponent $\varepsilon$ and thus test this possibility. To do so, we first track the defects between adjacent time samples, and calculate a finite difference velocity $\textbf{u}_k(t) = [\textbf{r}_k(t) - \textbf{r}_k(t~-~\Delta t)] / \Delta t$, where $\textbf{u}_k$ and $\textbf{r}_k$ are the velocity and position of the $k$th vortex, respectively, and $\Delta t$ is the sampling time step. This velocity is then averaged over all vortices in the system and all configurations in the ensemble to obtain $\umean(t)$. Although simplistic, this measurement appears to capture the overall behaviour of the mean vortex velocity. Expressed in relation to the vortex density, we expect $\umean \sim \nv ^ {(1 - \varepsilon) / 2}$ for a power-law mobility [using assumption (ii) above and taking $\nv \sim \Lc^{-2}$]. The PGPE $\nv(t)$ data from within the scaling window are shown in Fig.~\ref{fig:PVcomparison}(a), for which $z = 1.74(3)$, as stated in Sec.~\ref{sub:scaling}. The $\umean(\nv)$ data are presented in (b), with a power-law fit giving $\umean \sim \nv^{0.37(5)}$, i.e.~$\varepsilon = 0.27(9)$. The measurements of $z$ and $\varepsilon$ are in accordance with the prediction \eqref{eq:z<2}; hence we conclude that the exponent $z<2$ as measured in the PGPE is consistent with a power-law vortex mobility, with $\varepsilon > 0$. In Appendix~\ref{app:uv_dLdt}, we also demonstrate that assumption (i) is reasonably satisfied for this data set.

These results support a direct relation between $z$ and the vortex dynamics, but raise the question of the origin of the power-law mobility. To investigate further, we perform simulations of the aforementioned weakly damped point-vortex (PV) model. In this model, the velocity $\textbf{u}_k$ of vortex $k$ is given by:
\begin{equation} \label{eq:pointvortex}
    \textbf{u}_k = \left( 1 - \gamma_{\rm PV} s_k \hat{\textbf{z}} \, \times \, \right) \textbf{u}_k^{(0)},
\end{equation}
where
\begin{align}
    \textbf{u}_k^{(0)} = \frac{\pi \hbar}{m L} \sum_{j \neq k} s_j \sum_{q=-\infty}^{\infty} \Bigg( &\frac{\sin(2 \pi y_{jk} / L)}{\cosh[2\pi(x_{jk}/L - q)] - \cos(2\pi y_{jk} / L)}, \nonumber \\
    &\frac{ -\sin(2 \pi x_{jk} / L)}{\cosh[2\pi(y_{jk}/L - q)] - \cos(2\pi x_{jk} / L)}  \Bigg)
\end{align}
is the conservative equation of motion for a configuration of point-vortices at locations $\lbrace x_k, y_k \rbrace$ in a square domain of sidelength $L$ with periodic boundary conditions~\cite{weiss_nonergodicity_1991}. Here,  $s_k = \pm 1$ is the circulation sign of vortex $k$, $x_{jk} = x_j - x_k$ (likewise for $y_{jk}$), and $\gamma_{\rm PV} \ll 1$ is a phenomenological damping parameter that models the loss of energy to sound waves present in the PGPE. As initial conditions, we use the vortex positions extracted from our $\mathcal{N}=256$ PGPE simulations at $t = 3400 \, \hbar/\mu$, which is the beginning of the scaling window (on average, $\approx 130$ vortices remain at that time). We then solve the above equations using a semi-implicit integration scheme, with the inner sum truncated to $-3 \leq q \leq 3$~\cite{billam_onsager-kraichnan_2014}. We incorporate defect annihilation into the PV simulations by removing vortex--antivortex pairs if they come within $\xi$ of one another.

Under time evolution, the point-vortex density is seen to decay, as shown in Fig.~\ref{fig:PVcomparison}(a). We find that a choice of $\gamma_{\rm PV}=0.01$ results in almost immediate power-law scaling $\nv(t) \sim t^{-2/z}$, in agreement with the PGPE. The vortex configuration also remains similar to the PGPE throughout the evolution, as evidenced by the nearest-neighbour vortex correlations (see Appendix \ref{app:Cnn}). However, the exponent as measured from a power-law fit within the window $6000 \leq \mu t / \hbar \leq 10\,000$ gives $z=1.99(1)$, consistent with $z=2$ (other values of $\gamma_{\rm PV}$ delay the onset of power-law scaling, but eventually also result in $z \approx 2$). The mean point-vortex velocity $\umean$ is obtained by averaging Eq.~\eqref{eq:pointvortex} over all vortices, and the resulting $\umean(\nv)$ scaling is shown in Fig.~\ref{fig:PVcomparison}(b). A power-law fit gives $\varepsilon=0.04(4)$, in agreement with Eq.~\eqref{eq:z<2} for $z \approx 2$, and consistent with a constant vortex mobility as expected. The value $\varepsilon>0$ measured in the PGPE must therefore arise from a beyond-point-vortex effect, suggesting that it may result from vortex--sound interactions, which are absent in the PV model.

Turning to our Model A data, we are unable to reliably measure the mobility directly, because the slow drift velocity resulting from vortex interactions is overwhelmed by a fluctuating fast velocity arising from the noise. This issue also arises in the SPGPE for $\gamma \gtrsim 0.1$.

\subsection{Scaling of the dipole pair distribution}
In Refs.~\cite{chu_quenched_2001, forrester_exact_2013}, an analytic model was introduced for describing the relaxational dynamics of a two-dimensional superfluid following a temperature quench. In this description, the system is represented as a vortex-dipole pair distribution function $\Gamma(r,t)$, whose evolution is governed by a Fokker--Planck equation. The distribution $\Gamma(r,t)$ represents the probability of finding a vortex dipole of separation $r$ in the vortex configuration, and can be integrated to give the mean vortex density, $\nv(t) = 2 \int \Gamma(r,t) \diff^2 \textbf{r}$ (the prefactor here accounts for the two vortices per dipole). In Ref.~\cite{forrester_exact_2013}, it was predicted that the dipole distribution should obey a scaling form $\Gamma(r,t) \sim \Lc^{-\zeta} (t) F_{\rm d}(r / \Lc (t))$ with scaling function $F_{\rm d}$, assuming that $\Lc \sim t^{1/z}$. The exponent $\zeta$ is predicted to depend on the initial and final temperatures of the quench; but in particular, $\zeta=4$ for quenches from the Berezinskii--Kosterlitz--Thouless critical temperature.

\begin{figure}[b]
    \centering
    \includegraphics[width=\columnwidth]{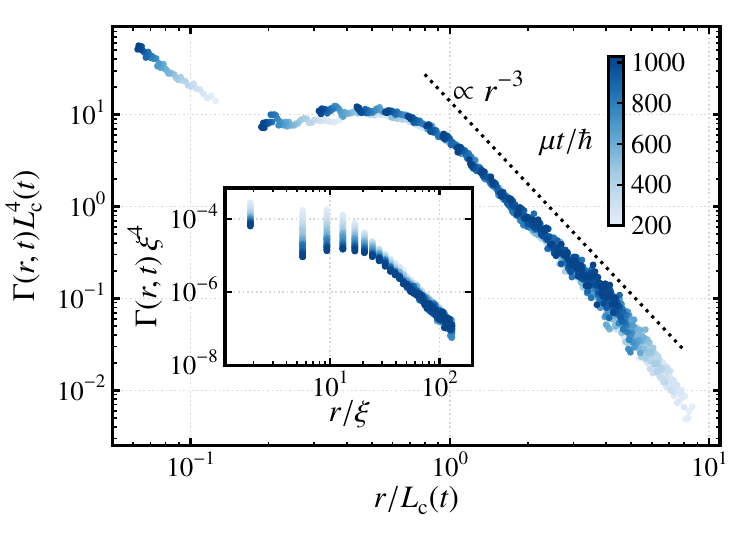}
    \caption{Evolution of the dipole pair distribution $\Gamma(r, t)$ for our Model A ($\gamma \to \infty$) simulation. The rescaled (raw) data are displayed in the main frame (inset), with the time indicated in each by the colorbar.}
    \label{fig:dipole_distribution}
\end{figure}

Given that relaxational dynamics are assumed in Refs.~\cite{chu_quenched_2001, forrester_exact_2013}, our Model A simulations are the closest point of comparison to the predictions stated above. Although there is an inherent ambiguity in assigning a configuration of vortices into a set of dipoles, we nonetheless attempt a measurement of $\Gamma(r,t)$ using the same prescription as in Ref.~\cite{jelic_quench_2011}. We first rank in increasing order the distance between every possible pair of opposite circulation vortices in the system (taking into account the periodic boundary conditions). Beginning with the smallest pair, we proceed sequentially through this list, assigning a pair as a dipole only if neither of its constituent vortices has already been assigned to another dipole. In this way, we assemble a unique list of $N_{\rm v}/2$ pairs, where every vortex has been paired exactly once ($N_{\rm v}$ is the total number of vortices in the system). We then construct a histogram $n_{\rm d}(r,t)$ of all dipole sizes $r$ in the system, and average the distribution over all simulations in the ensemble. Finally, we define $\Gamma(r,t) = n_{\rm d}(r,t) / (2 \pi r L^2)$; here, the factor of $L^2$ converts to a spatial density, while the factor of $2 \pi r$ accounts for the number of ways of configuring a dipole of size $r$.

Figure~\ref{fig:dipole_distribution} shows the resulting distribution $\Gamma(r,t)$ for times within our identified scaling window (see Sec.~\ref{sec:analysis}). Upon rescaling the data according to the above prediction with $\zeta=4$, a convincing collapse is obtained. We also observe a power-law tail in the dipole distribution, $\Gamma(r,t) \sim r^{-3}$, for $r \gtrsim \Lc$. The peak visible in the smallest radial bin can be attributed to thermal dipoles~\cite{simula_thermal_2006} that appear throughout the system during the dynamics. We conclude that our Model A data is consistent with this vortex-dipole description of the coarsening process.

\section{Conclusions}

We investigated the coarsening of a scalar 2D Bose gas following an instantaneous quench into the ordered phase. By varying the dimensionless dissipation rate $\gamma$, we explored the crossover between purely relaxational (Model A, $\gamma\to\infty$) and conservative ($\gamma=0$) dynamics. Our results for Model A are consistent with dynamical critical exponent $z=2$ with logarithmic corrections, the generally agreed result in the literature. Our central result is that for decreasing dissipation rate $\gamma$ we continue to observe universal scaling in time, but with a smooth reduction in the exponent towards a value $z < 2$ in the conservative limit. We found evidence that the deviation from $z=2$ may be attributed to an anomalous power-law vortex mobility that arises from interactions between vortices and sound waves. For Bose gas experiments, $\gamma \lesssim 0.02$ is typical~\cite{weiler_spontaneous_2008, rooney_persistent-current_2013, ota_collisionless_2018, liu_dynamical_2018}, suggesting that this anomalous behaviour is likely to be observable. Sudden quenches into the ordered phase were recently implemented in a homogeneous three-dimensional Bose gas experiment \cite{glidden_bidirectional_2021}; similar quenches could soon be possible in quasi-2D setups. In future, it will be interesting to further investigate the source of the power-law vortex mobility identified here.

Data supporting this publication are openly available under a Creative Commons CC-BY-4.0 License found in Ref.~\cite{data}.

\begin{acknowledgments}
We thank Tapio Simula for suggesting a power-law function for the vortex mobility, and acknowledge useful discussions with Carlo Barenghi and Franco Dalfovo, as well as financial support from the UK EPSRC [Grant No. EP/R021074/1] (AJG and TPB) and the Quantera ERA-NET cofund project NAQUAS [EPSRC Grant No. EP/R043434/1] (PC and NPP). This research made use of the Rocket High Performance Computing service at Newcastle University.
\end{acknowledgments}

\appendix
\renewcommand{\thefigure}{A\arabic{figure}}
\setcounter{figure}{0}

\section{Technical details of the PGPE initial condition \label{app:ICs}}

For the $\gamma=0$ (PGPE) simulations, a careful choice of initial condition must be made, because of the constraints of conservation of energy and particle number (as described in Sec.~\ref{sec:simulations}). To facilitate direct comparison between the $\gamma=0$ and $\gamma>0$ cases, the mean energy- and particle-densities for the $\gamma=0$ initial states are chosen to be equal to their ensemble-averaged values, $\bar{\epsilon} = \bar{E}/L^2$ and $\bar{n} = \bar{N}/L^2$, in the $\gamma>0$ system after equilibration (i.e.,~at $t \to \infty$). This is achieved by initiating the wavefunction as a populated disk of radius $k_{\rm d}$ in wavenumber space, $\psi = \sum_{|\textbf{k}| < k_{\rm d}} \sqrt{n_\textbf{k}} \exp{[i (\textbf{k} \cdot \textbf{r} + \phi_\textbf{k})]}$. The populations $n_\textbf{k}$ are chosen to be uniform and to ensure the correct mean density $\bar{n}$. The phase $\phi_\textbf{k}$ of each mode is initially randomised, and a Powell minimisation algorithm~\cite{powell_efficient_1964} is subsequently used to adjust the phases to achieve mean energy-density $\bar{\epsilon}$. The radius of the disk is set to $k_{\rm d}= 0.2 \, k_{\rm cut}$.

\begin{figure}[b]
    \centering
    \includegraphics[width=\columnwidth]{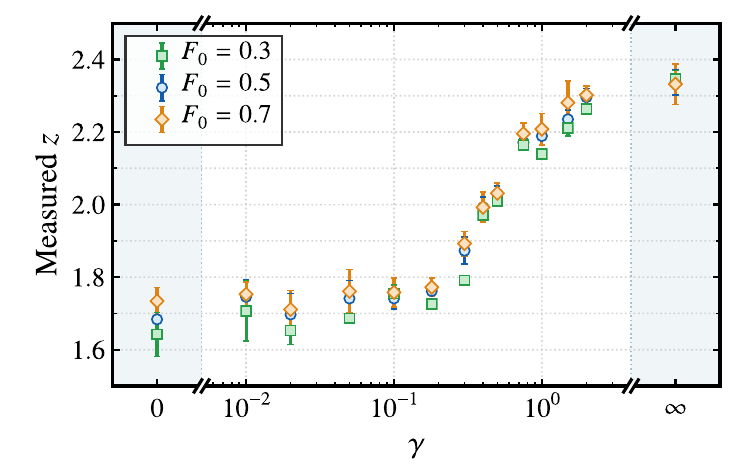}
    \caption{Critical exponent $z$, measured from power-law fits to $\Lc(t)$, as a function of dimensionless dissipation rate $\gamma$. Here, $\Lc(t)$ has been extracted using three different scaling function thresholds $F_0$.}
    \label{fig:zvsgamma_threshold}
\end{figure}

\begin{figure}[t]
    \centering
    \includegraphics[width=\columnwidth]{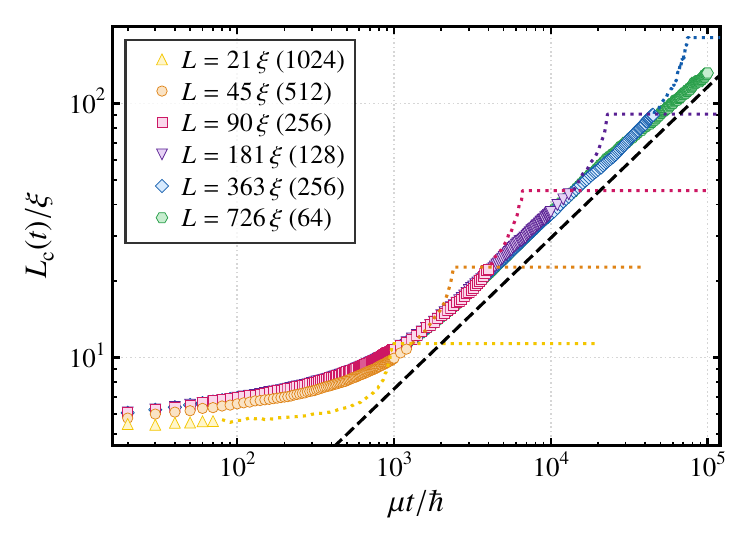}
    \caption{Evolution of $\Lc(t)$ in various system sizes $L$ for $\gamma = 0$ (the PGPE). To give an indication of where finite size effects begin to dominate, the data points are replaced by dotted lines of the corresponding colour at all times for which $\Lc(t)>L/4$. The power-law fit to the scaling window of the $L = 363 \, \xi$ case is shown as a black dashed line. The plateau at late times corresponds to the maximum possible correlation length of $\Lc = L/2$. The ensemble size $\mathcal{N}$ for each system is given in brackets in the legend.}
    \label{fig:Lc_vs_L}
\end{figure}

\begin{figure}[t]
    \centering
    \includegraphics[width=\columnwidth]{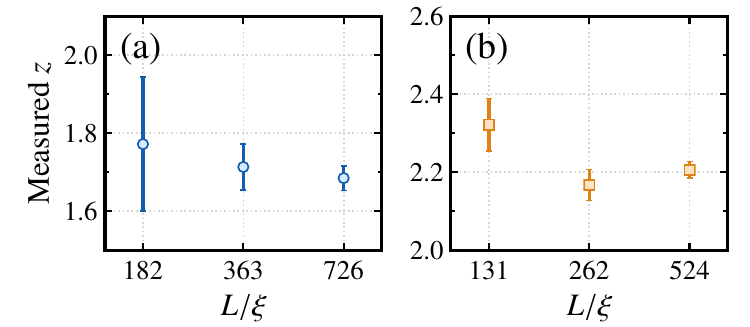}
    \caption{The measured value of $z$ from fits to $\Lc(t)$ in different system sizes $L$ for (a) $\gamma=0$ and (b) $\gamma=1$.}
    \label{fig:z_vs_L}
\end{figure}

\section{Dynamical movies \label{app:movies}}
Included in the Supplemental Material are movies of the classical field dynamics for $\gamma= \lbrace 0, 0.1, 1, \infty \rbrace$ [the last of these is achieved by setting $\alpha=0$, $\gamma=1$ in Eq.~\eqref{eq:SPGPE}]. From left to right, the panels in these movies correspond to the classical field density $|\psi(\textbf{r},t)|^2$, the phase $\mathrm{arg} \left \lbrace \psi(\textbf{r},t) \right \rbrace$, and the locations of vortices and antivortices, obtained by numerically identifying all phase windings of $\pm 2\pi$ in the field. At each time, the number of vortices and antivortices is shown in the lower left of the leftmost panel, and the \textit{physical time} is shown in the upper left. For brevity, the rate at which physical time passes in the movies increases at the beginning of each decade in physical time (when this happens, the movie briefly pauses). For the $\gamma>0$ movies, the $\psi=0$ initial condition is used, and the mean density is seen to grow rapidly, plateauing by $t \sim 100 \ \hbar/\mu$. Note that there is no input temperature parameter $T$ for the $\gamma=0$ simulation, but we have chosen the effective equilibrium temperature to be the same as for $\gamma>0$ by restricting the number- and energy-density (details of the $\gamma=0$ initial condition are provided in Appendix~\ref{app:ICs}). In all movies, tightly bound thermal dipoles~\cite{simula_thermal_2006} are seen to spontaneously appear in the superfluid from time to time, surviving only briefly before annihilating again.

\begin{figure}[t]
    \centering
    \includegraphics[width=\columnwidth]{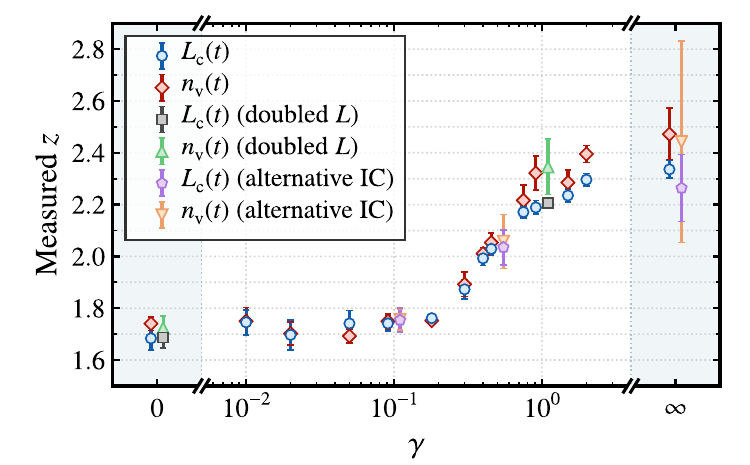}
    \caption{Measured critical exponent $z$ as a function of dimensionless dissipation rate $\gamma$. The blue circles and red diamonds are the same as in Fig.~\ref{fig:z_vs_gamma}, while the additional data correspond to quenches at $\gamma=\lbrace 0, 1 \rbrace$ with a doubled system size $L$, and multiple quenches at $\gamma > 0$ using our alternative nonzero density initial condition (IC). Where multiple sets of data fall onto a single $\gamma$ value, the points are symmetrically offset along the horizontal axis for clarity.}
    \label{fig:zvsgamma_extradata}
\end{figure}

\section{Possible systematic effects \label{app:systematics}}

\emph{Choice of threshold}---We find that the choice of scaling function threshold $F_0$ (used for defining the correlation length) has a slight systematic effect on the power-law scaling of $\Lc(t)$. In Fig.~\ref{fig:zvsgamma_threshold}, the critical exponent $z$ measured from power-law fits to $\Lc(t)$ is shown as a function of dissipation rate $\gamma$, with the correlation length extracted using three different choices of $F_0$. The fitting is carried out in the same way as described in Sec.~\ref{sub:scaling}. Evidently, a larger $F_0$ gives rise to a slightly larger measured exponent $z$ (on average). As noted in Sec.~\ref{sec:analysis}, $F_0$ should be chosen to minimise both discretisation effects at small scales and finite size effects at large scales. As such, we do not explore thresholds outside of $0.3 \leq F_0 \leq 0.7$ here.

\emph{Lack of finite size effects}---It is expected that an infinitely large system undergoing dynamical scaling should exhibit power-law growth of $\Lc(t)$ indefinitely~\cite{bray_theory_1994}. In a finite system of size $L$, on the other hand, scaling must eventually cease once $\Lc(t)$ grows to $\sim L$. However, with increasing $L$, we expect power-law growth to continue for increasingly long times before finite size effects dominate. To demonstrate that this is the case in our simulations, we repeat our $\gamma=0$ quench at six values of $L$ (with fixed grid spacing $\approx 0.7\,\xi$). Figure~\ref{fig:Lc_vs_L} shows the resulting $\Lc(t)$ evolution. To indicate the time at which finite size effects begin to strongly affect the correlation length for each $L$, we show the data as dashed lines for $\Lc(t) > L/4$ (note that this condition is less stringent than that described in Sec.~\ref{sec:analysis} for choosing the end of our fitting windows). We find that the scaling regime is only convincingly reached in the largest three system sizes shown here.

\begin{figure*}[t]
    \centering
    \includegraphics[width=2\columnwidth]{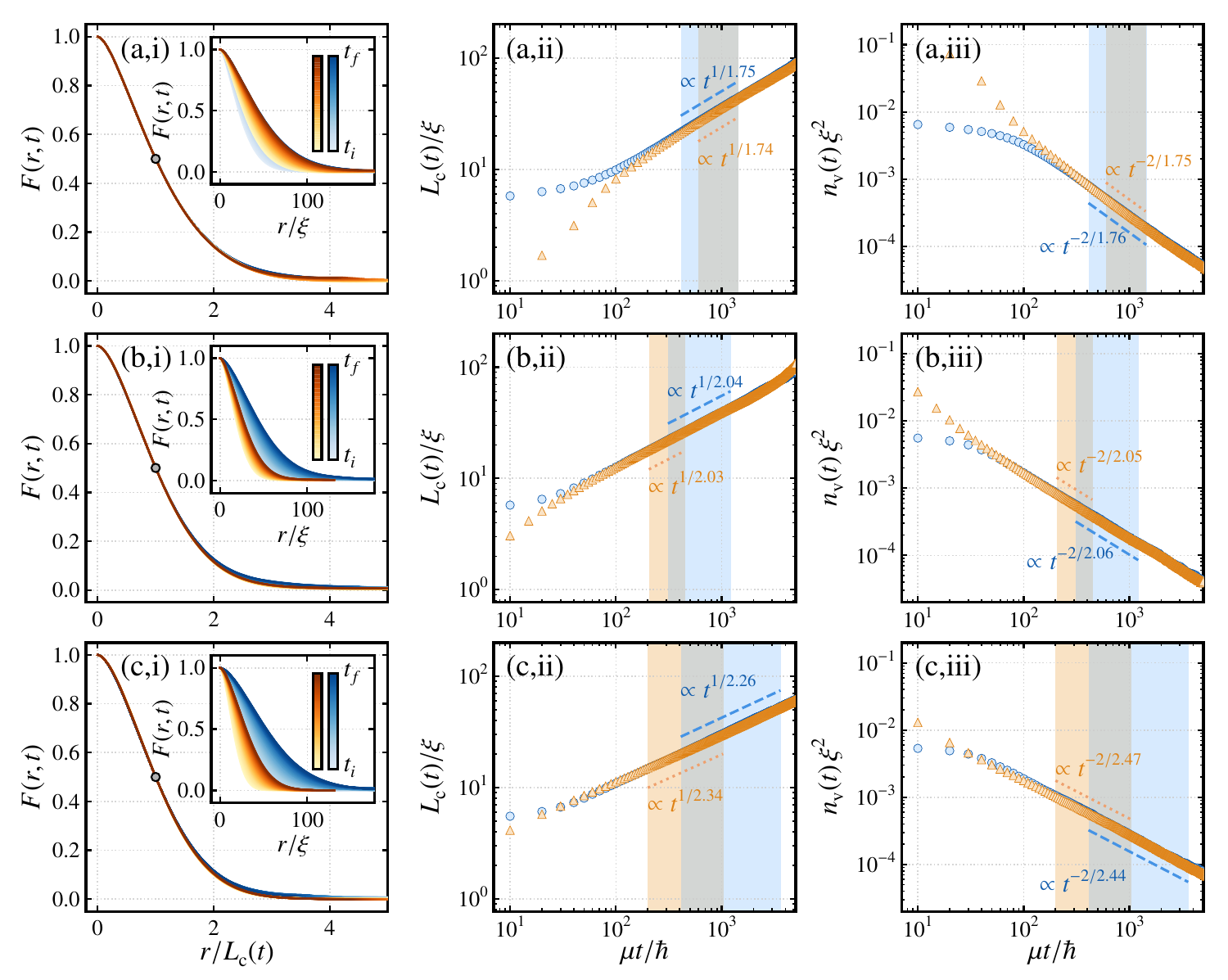}
    \caption{Comparison of scaling behaviour between the zero (orange/triangles) and nonzero (blue/circles) density initial conditions, for dissipation rates $\gamma=0.1$ (a), $\gamma=0.5$ (b), and $\gamma \to \infty$ (c). Ensemble averaged scaling function $F(r,t)$ plotted against radial distance $r$ (i), both before (inset) and after (main frame) rescaling by the correlation length $\Lc(t)$. The time at which each curve has been sampled is denoted by the color bar in the insets, and a grey dot signifies the threshold $F_0=0.5$ used to define $\Lc$. Evolution of the mean correlation length $\Lc(t)$ (ii) and vortex density $\nv(t)$ (iii). In columns (ii,iii), the scaling window is highlighted in the appropriate colour, and the best power-law fit to the data is shown as an orange dotted (blue dashed) line for the zero (nonzero) density initial condition (offset for clarity). The initial and final times of the scaling window, $t_i$ and $t_f$, can be read off the horizontal axis in columns (ii,iii).}
    \label{fig:scaling_comparison}
\end{figure*}

As $L$ is increased, we also expect that the exponent $z$ as measured from a power-law fit to $\Lc(t)$ should display convergence of the best-fit value, as well as reduced uncertainty. We show that this is the case in Fig.~\ref{fig:z_vs_L} by comparing $z$ as measured from three $L$ values with both $\gamma=0$ [panel (a)] and $\gamma=1$ [panel (b)]. The central $L$ values here are the same as those defined in Sec.~\ref{sec:simulations}. To determine $z$ in each case, we apply the same fitting technique as described in the Sec.~\ref{sub:scaling}. Since the duration of the scaling window increases with increasing $L$ (as evidenced by Fig.~\ref{fig:Lc_vs_L}), for consistency we use a minimum fitting window of $N_{\rm s}/4$ points when constructing the $z$ histogram. Here, $N_{\rm s}$ is the number of sampled points within the scaling window. We also fix the ensemble size to $\mathcal{N}=64$ for these measurements. In Fig.~\ref{fig:z_vs_L}, the error in the $z$ measurement is seen to decrease with $L$, as expected. For $\gamma=1$, the $z$ value measured in the $L=131 \, \xi$ system does not overlap with the values from the larger systems, suggesting that this system is not sufficiently large to obtain an accurate measurement of $z$.

For comparison with all of our $z(\gamma)$ data, we include the $z$ values measured from the doubled system sizes at $\gamma=\lbrace 0, 1 \rbrace$ in Fig.~\ref{fig:zvsgamma_extradata}, alongside the data from Fig.~\ref{fig:z_vs_gamma}.

\emph{Lack of initial condition dependence}---In the theory of phase ordering kinetics~\cite{bray_theory_1994}, it is generally expected that the precise form of initial conditions used to initiate the coarsening behaviour should play no role in determining the observed scaling. To confirm that this holds for our system, we have repeated our $\gamma= \lbrace 0.1, 0.5, \infty \rbrace $ simulations using the nonzero density initial condition that was used for the $\gamma=0$ simulations (see description in previous section). We do this using a system size of $L \approx 363 \, \xi$, and ensembles of $\mathcal{N}=64$ ($\mathcal{N}=256$) trajectories for $\gamma = \lbrace 0.5, \infty \rbrace$ ($\gamma=0.1$). In Fig.~\ref{fig:scaling_comparison}, we compare the evolution of the scaling function [column (i)], correlation length [column (ii)] and vortex density [column (iii)] for the two initial conditions, with $\gamma=\lbrace 0.1, 0.5, \infty \rbrace$ in rows (a,b,c), respectively. The procedures used to identify the scaling windows and perform fits to the data have been carried out as described in Secs.~\ref{sec:analysis} and \ref{sub:scaling}. For all $\gamma$ values, the rescaled $F(r,t)$ data [main frames of column (i)] are almost indistinguishable between the two initial conditions. Likewise, the observed power-law scaling in columns (ii) and (iii) appears to be almost unaffected by the initial condition, despite substantial differences in the curves at early times. We note that for the zero (nonzero) initial density configurations, the curves approach the scaling law from steeper (shallower) evolution. This provides strong evidence that the system has reached a universal scaling regime.

Figure~\ref{fig:zvsgamma_extradata} displays the critical exponents $z$ measured from fits to both $\Lc(t)$ and $\nv(t)$ in these simulations. In all cases, these exponents are consistent (within our estimated uncertainties) with the exponents measured for the same $\gamma$ value using the zero density initial condition.

\begin{figure}[b]
    \centering
    \includegraphics[width=\columnwidth]{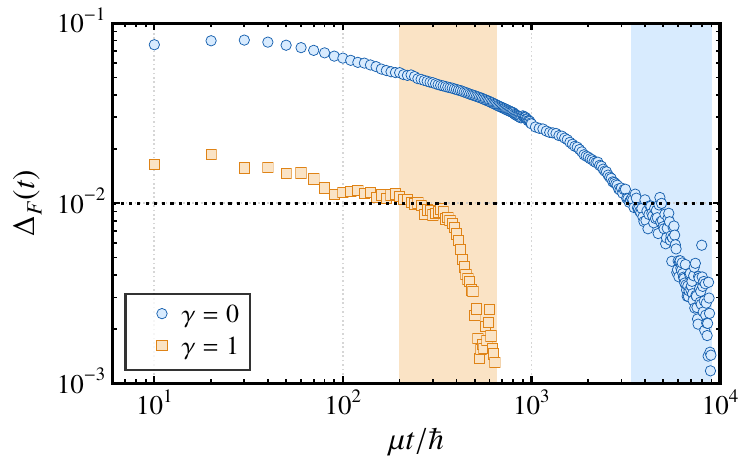}
    \caption{Evolution of $\Delta_F(t)$ for $\gamma= \lbrace 0, 1 \rbrace$. The scaling window for $\gamma=0$ ($\gamma=1$) is denoted by the right (left) highlighted region. The dotted line denotes the threshold value of $\Delta_F = 0.01$.}
    \label{fig:DeltaF}
\end{figure}

\section{Quantification of the scaling function collapse \label{app:DeltaF}}
To quantify the precision of the collapse of the scaling function $F(r,t)$, we define the metric:
\begin{equation}
    \Delta_F(t) = \mathrm{max}_r \left \lbrace \left | F(r/\Lc(t), t) - F(r/\Lc(t^\prime), t^\prime) \right | \right \rbrace
\end{equation}
where $t^\prime$ is a reference time, which we choose to be the end of the scaling window (as defined in Sec.~\ref{sec:analysis}). A value of $\Delta_F \ll 1$ indicates good agreement between the measured scaling functions at the two times $t$ and $t^\prime$.

In Fig.~\ref{fig:DeltaF}, we plot $\Delta_F(t)$ for $\gamma= \lbrace 0, 1 \rbrace$ (for comparison with Fig.~\ref{fig:scaling_data}). From this data, it can be seen that the beginning of the scaling windows as chosen by eye (highlighted regions) correspond to $\Delta_F (t) \approx 0.01$. Our windows therefore comprise collapses that are accurate to within a maximum deviation of $\approx 1\%$.

\section{Analysis of the vortex configuration \label{app:Cnn}}
As an indicator of the vortex configuration at a given time, we calculate the nearest-neighbour correlator,
\begin{equation}
C_{\rm nn} = \frac{1}{N_{\rm v}} \sum_k^{N_{\rm v}} s_k s_k^{\rm nn},
\end{equation}
where $N_{\rm v}$ is the total number of vortices, $s_k = \pm 1$ is the circulation sign of vortex $k$, and $s_k^{\rm nn}$ is the circulation sign of its nearest neighbour. A value of $-1 \leq C_{\rm nn} \lesssim 0$ indicates a vortex configuration predominantly paired into dipoles, $0 \lesssim C_{\rm nn} \leq 1$ indicates clustering of same-sign vortices, and $C_{\rm nn} \approx 0$ corresponds to an approximately random vortex distribution~\cite{billam_onsager-kraichnan_2014}.

\begin{figure}[t]
    \centering
    \includegraphics[width=\columnwidth]{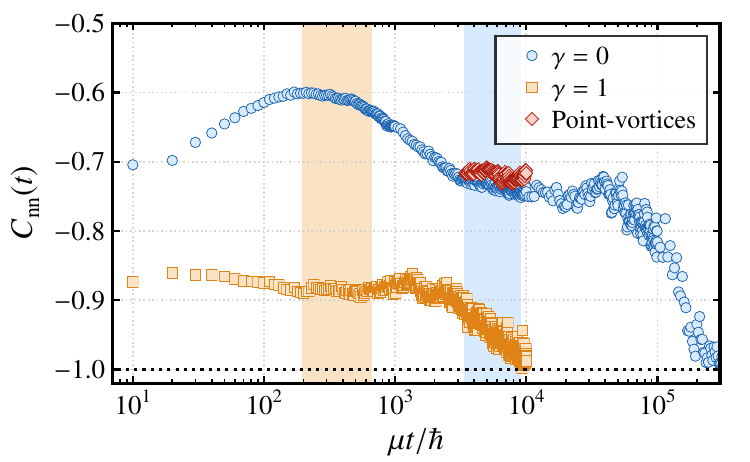}
    \caption{Nearest-neighbour vortex correlator $C_{\rm nn}(t)$ for $\alpha=1$, and $\gamma= \lbrace 0, 1 \rbrace$, as in Fig.~\ref{fig:scaling_data}. The scaling window for $\gamma=0$ ($\gamma=1$) is denoted by the right (left) highlighted region. Also shown is the correlator measured from our point-vortex simulations (see Sec.~\ref{sub:zl2}). The dotted line corresponds to the minimum possible value of $-1$.}
    \label{fig:Cnn}
\end{figure}

Figure~\ref{fig:Cnn} shows the evolution of the mean $C_{\rm nn}(t)$ as measured from the $\gamma=0$ and $\gamma=1$ simulations (for comparison with Fig.~\ref{fig:scaling_data}). Throughout the evolution, $C_{\rm nn}<0$ in both cases, indicating that the vortices are in the dipolar regime. Interestingly, the vortices are distributed quite differently within the respective scaling windows, as evidenced by the significantly different values of $C_{\rm nn}$.

\section{Relationship between mean vortex velocity and correlation length \label{app:uv_dLdt}}
In Sec.~\ref{sub:zl2} we assume that in the $\gamma=0$ system the mean vortex velocity and the correlation length are related via
\begin{equation} \label{eq:dLdt}
    \frac{\diff \Lc }{\diff t} = A \bar{\rm u}_{\rm v}(t)
\end{equation}
for some dimensionless factor $A$. Here we provide evidence from our $\gamma=0$ simulations to support this relationship.

\begin{figure}[t]
    \centering
    \includegraphics[width=\columnwidth]{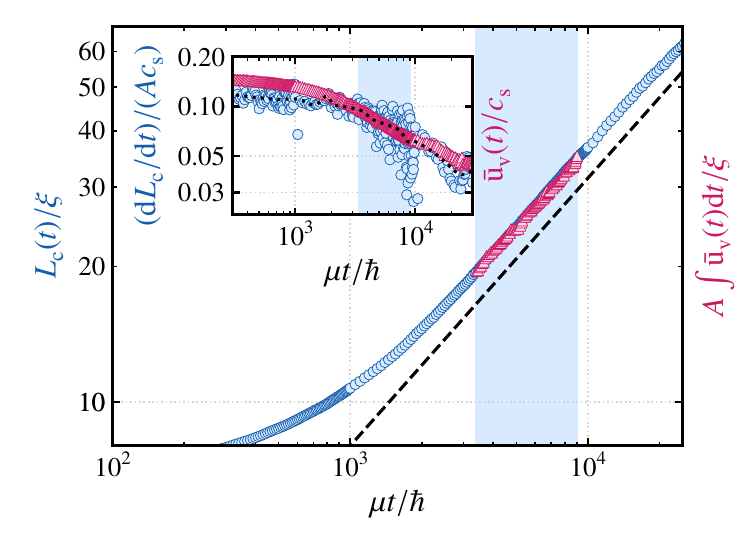}
    \caption{Comparison between the two sides of Eq.~\eqref{eq:dLdt} for our $\gamma=0$ simulations. Blue circles show the correlation length (main frame) and its numerical derivative (inset), while pink triangles show the mean vortex velocity (inset) and its integral (main frame). The black dashed line in the main frame shows the power-law fit to $\Lc(t)$, and the highlighted region identifies the scaling window [as in Fig.~\ref{fig:scaling_data}(a,ii)]. The black dotted line in the inset shows a smoothed spline fit to the numerically differentiated $\Lc(t)$ data. In both panels, $A=0.036$ is used to give the best agreement within the scaling window.}
    \label{fig:dLdt_vs_uv}
\end{figure}

We first numerically differentiate $\Lc(t)$ (using a second-order central difference method) to obtain the left-hand-side of Eq.~\eqref{eq:dLdt}; the result is shown in the inset of Fig.~\ref{fig:dLdt_vs_uv} (blue circles). To reduce the effect of noise arising from numerical differentiation, we also show a smoothed spline fit to the data (dotted line). For comparison, we plot on the same axis $\umean(t)$ as measured from our vortex tracking (pink triangles). The two datasets are seen to broadly agree for all times shown.

As a secondary method of affirming Eq.~\eqref{eq:dLdt}, we integrate both sides to give $\Lc(t) = A \int \umean(t) \diff t$. Assuming $\umean(t) \sim t^\beta$ within the scaling window, the integral on the right-hand-side of this equation yields $\int \umean(t) \diff t = t \umean(t) / (1 + \beta)$. In the main frame of Fig.~\ref{fig:dLdt_vs_uv}, we plot $\Lc(t)$ [blue circles; same data as in Fig.~\ref{fig:scaling_data}(a,ii)], alongside the above form of the integral $\int \umean(t) \diff t$ (pink triangles). While the integration method only applies within the scaling window, it provides a cleaner comparison than numerical differentiation.

With both of the above methods, we find the proportionality factor in Eq.~\eqref{eq:dLdt} to be $A \approx 0.04$ within the scaling window. This factor should quantify how much of the mean vortex velocity contributes to the rate of change of the correlation length. It may therefore be loosely interpreted as the effective dissipation rate for the $\gamma=0$ system that arises from loss of energy to sound waves. However, we note that the precise value of $A$ is susceptible to systematic shifts based on the manner in which both $\Lc(t)$ and $\umean(t)$ are measured.

%

\end{document}